\documentclass{PoS}

\title{Detecting extended gamma-ray emission with the next generation Cherenkov telescopes}

\ShortTitle{Detecting extended emission from blazars}

\author {M. Fernandez Alonso$^a$, A.D. Supanitsky$^a$, \speaker {A.C. Rovero}$^a$\\
\llap{$^a$}Instituto de Astronom\'ia y F\'isica del Espacio (IAFE, CONICET-UBA), Buenos Aires, Argentina.\\
E-mail: \email{mateofa@iafe.uba.ar}} %,

\abstract{Very high energy (VHE $>$100 GeV) gamma rays coming from blazars can produce pairs when interacting with the Extragalactic Background Light (EBL) and the Cosmic Microwave Background, generating an electromagnetic cascade. Depending on the Intergalactic Magnetic 
Field (IGMF) intensity, this cascade may result in an extended isotropic emission of photons around the source ({\it halo}), 
or in a broadening of the emission beam. The detection of these effects might lead to important constrains both on the IGMF intensity and the EBL density, quantities of great relevance in cosmological models. Using a Monte Carlo program, we simulate 
electromagnetic cascades for different values of the IGMF intensities and coming from a source similar to 1ES0229+200, a blazar with hard intrinsic spectrum at redshift $z=0.14$, which is an ideal distance for potentially observing the effect. We study the possible response of a generic future Cherenkov telescope using a simplified model for the sensitivity, effective area and angular resolution. Finally, combining these instrument properties, we calculate the angular distribution of photons and develop a method to test the statistical feasibility of detecting the effect in the near future.}


\FullConference{The 34th International Cosmic Ray Conference,\\
		30 July- 6 August, 2015\\
		The Hague, The Netherlands}

\begin{document}

\section{Introduction}

The Universe is opaque for gamma rays in the VHE ($>$100 GeV) range. Photon absorption in the intergalactic (IG)  photon backgrounds is energy dependent and starts to become substantial at TeV energies \cite{gould1966}. In particular, VHE gamma rays from jets of active galactic nuclei (AGN) can interact with photons in the IR-UV range present in the Extragalactic Background Light (EBL) and photons from the Cosmic Microwave Background (CMB), producing electron-positron pairs. These pairs carry most of the energy from the original photons, and can interact as well with IG photons via Inverse Compton, promoting them to energies in the HE ($>$100 MeV)-VHE range, making them able to pair produce in the IG backgrounds again. This cascade process converts the initial VHE photons into photons of lower energy which can travel further. Moreover, depending on the intensity (B) of the Intergalactic Magnetic Field (IGMF), the bending effect on the electron-positron pair trajectories can result into different emission scenarios. For a {\it strong} IGMF intensity (B $>10^{-7}$ G) synchrotron cooling becomes dominant and no secondary gamma rays are produced \cite{gould1978}. For a {\it moderate} IGMF ($10^{-12}$ G $<$ B $<10^{-7}$ G) the electron and positron pair trajectories are isotropised around the source eventually giving rise to an extended isotropic emission of photons around the source, or {\it halo}, which take much longer to reach the observer than the direct photons from the source \cite{aha1994}. For a {\it weak} IGMF (B $<10^{-14}$ G) the cascade develops almost exclusively in the forward direction, although there is a broadening of the original emission beam, even for very small IGMF intensities. The extension of this emission depends on the IGMF intensity and should be clearly distinguished from the halo emission because in this case the broadening takes place along the jet direction, not in an isotropic way \cite{hess2014}. Since this effect was proposed several groups have tried unsuccessfully to observe it using multiple methods: \cite{aha2001}, \cite{alec2010}, \cite{fallon2010}, \cite{fernandez2014} and \cite{hess2014}. All this studies were done using blazars, a subtype of AGN that have their jets pointing towards the Earth and are therefore extremely luminous objects, perfect candidates to perform this type of studies. The detection of the halo, or the magnetic broadening might lead to important constrains on both the IGMF intensity and the EBL density, quantities that currently remain uncertain and are of great importance in cosmological models.\\
In this work we use MC based simulations of intergalactic cascades under different IGMF scenarios, and study the effects of the magnetic field on the resulting spectral and angular distributions of the arriving photons. Motivated by the existence of a real future Cherenkov telescope system, the Cherenkov Telescope Array \cite{CTA2013}, we assume a simplified model of response for such a type of system to develop a method for testing the feasibility of detection of extended emission around a given source.

\section{Simulation}
\label{sim}

We use the ELMAG program \cite{elmag2012} to simulate the intergalactic electromagnetic cascades. This simulation takes into account pair production, inverse Compton, synchrotron losses, and deflections of the charged particles in the IGMF. We consider a source similar to 1ES0229+200, an HBL type blazar at redshift $z=0.14$, that seems to have a particularly hard intrinsic spectrum, an important feature for the extended emission effect. We inject photons with energy spectrum following a power law with index $\Gamma=2/3$ \cite{dolag2010}, and adjust the VHE part of the flux to match current observations of this blazar \cite{stecker2008}. Here we do not try to explain the SED of this particular source, e.g. we are not contemplating Fermi-LAT observations at lower energies, neither considering other values for the spectral index. The mention of the blazar 1ES0229+200 is just to justify the existence of such a spectrum.

\begin{figure}[!ht]
  \centering
  \includegraphics[width=0.9\textwidth]{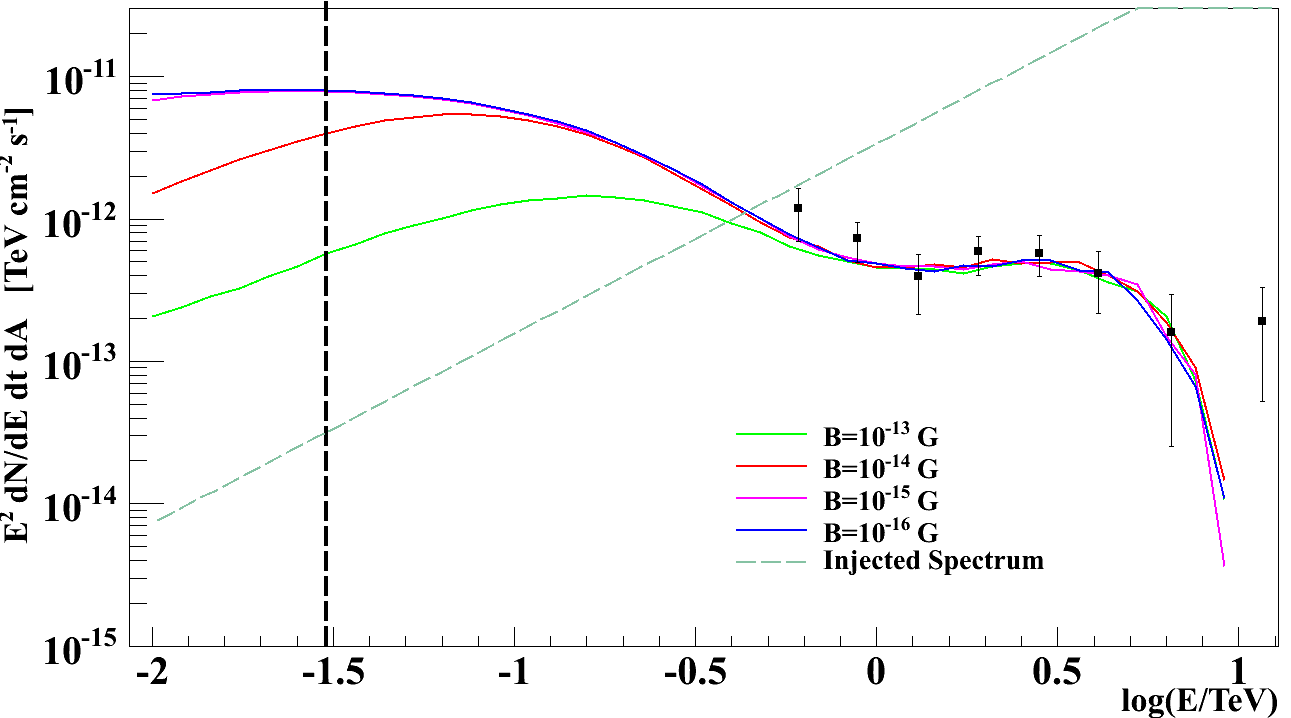}
  \caption{Spectral energy distribution of photons arriving at Earth as generated by the simulation and for different IGMF intensities. The SEDs are scaled to match the VHE part of the measured flux from 1ES0229+200 (black points).}
  \label{Simulation}
\end{figure}

We study the cascade considering different IGMF intensities: $B=10^{-13}$ G, $B=10^{-14}$ G,  $B=10^{-15}$ G and $B=10^{-16}$ G, which cover the range of interest for this effect \cite{hess2014}. We also consider the emission to be collimated in a cone of 6$^\circ$ \cite{dolag2011}, which is a more realistic treatment of blazar emission than previous approaches. Using this simulations we calculate the SED and the angular distribution ($\theta^2$) of photons arriving at Earth in the energy range 30 GeV - 10 TeV, which is the approximate energy window of future Cherenkov telescopes (see Figure \ref{Simulation}).

\vspace{.5cm}
\vspace{.5cm}

It is interesting to note the differences between the SEDs shown in Figure \ref{Simulation}, particularly for energies close 
to 30 GeV (dashed line), where the number of photons increases as the IGMF intensity weakens. This is a direct result of the deflection of pairs that end up scattering photons in directions off the source-earth line. Differences in the SED can be used to set limits to the IGMF (\cite{dolag2010}, \cite{taylor2011}). For an intensity of $B=10^{-16}$ G or weaker, radiation is collimated enough to make SEDs indistinguishable. Angular distributions however, present a broadening effect even at this weak field regime. Our study focuses on the possible broadening effect that a relatively weak IGMF may have in the $\theta^2$ distributions, and how this effect can be used to set limits to the IGMF when the SEDs show no significant differences. 
Following this idea, and considering that for weaker magnetic fields SEDs will present minimal differences, we continue our analysis for the case of $B=10^{-16}$ G.

\section{Telescope Response}
\label{tel}

\begin{figure}[!ht]
  \centering
  \includegraphics[width=0.9 \textwidth]{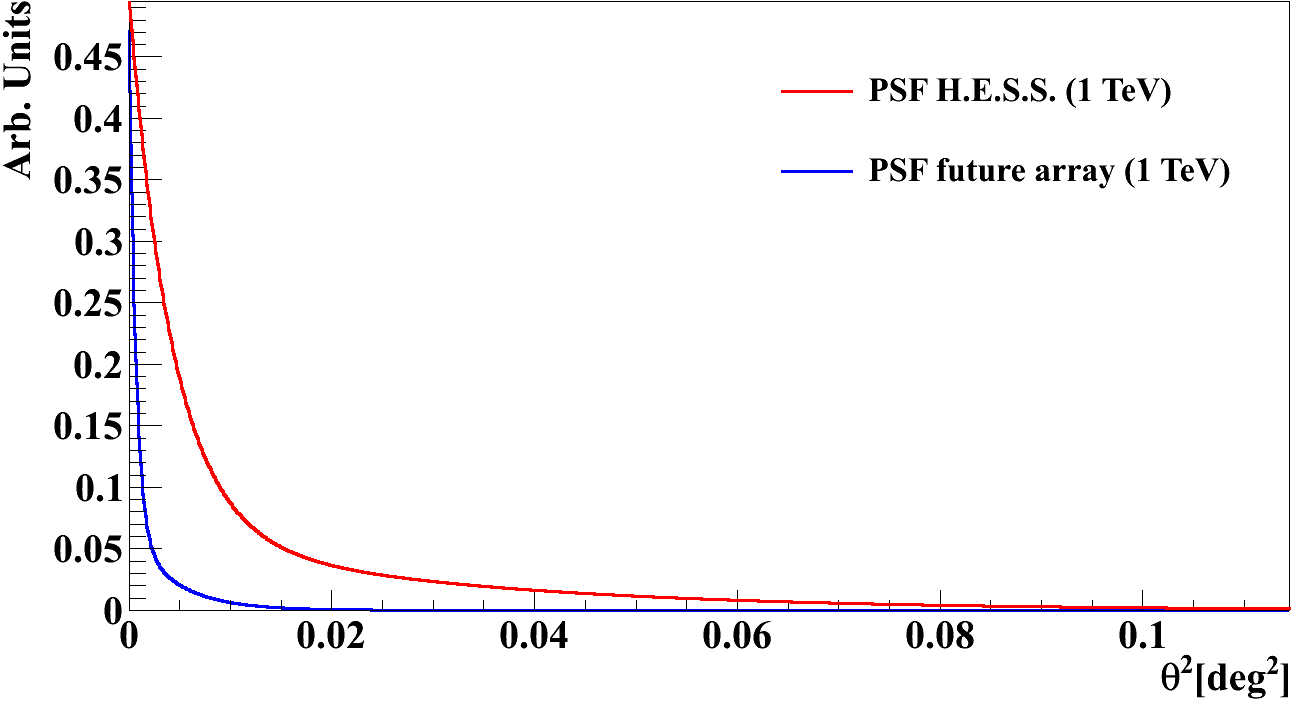}
  \caption{PSF for H.E.S.S. (red) and the estimated PSF for a future array (blue) for $E_{\gamma}=1$ TeV.}
  \label{PSF}
\end{figure}

As mentioned before, there have been several attempts to observe the effect, in all cases only upper limits could be obtained. Detection was not possible because the effect is either non existent, or too weak to be detected with present instruments. Future generation Cherenkov telescopes will enhance their sensitivity and angular resolution improving chances of detection. In this study we adopt the generic performance curves proposed by \cite{charbonnier2011}:
\begin{equation}
S = -13.1 - 0.33\ X + 0.72\ X^2
\label{sens}
\end{equation}
\begin{equation}
A = 6 + 0.46\ X - 0.56\ X^2
\label{area}
\end{equation}
\begin{equation}
\psi_{68} = 0.038 + \exp\left[ -(X + 2.9)/0.61\right] 
\label{psi}
\end{equation}
\\
where $X = \log(\textrm{Photon Energy/TeV})$, $S =$ Differential Sensitivity/(erg cm$^{-2}$ s$^{-1}$), $A =$ Effective Area/m$^2$ and $\psi_{68}$ is the 68\% containment radius of the point spread function (PSF) in
degrees. We also use the PSF shape obtained by H.E.S.S. \cite{aha2006},
\\
\begin{equation}
 PSF = C\left[ \exp\left( \frac{-\theta^{2}}{2\sigma_{1}^2}\right) +C_{rel}\ \exp\left( \frac{-\theta^{2}}{2\sigma_{2}^2}\right) \right] 
 \label{HESS}
 \end{equation}
\\
where $C_{rel}=0.15$ and $C$ is a global normalization factor. We combine this PSF shape with $\psi_{68}$ from equation 
(\ref{psi}) to construct the PSF corresponding to a future Cherenkov telescope. The resulting PSF is energy dependent and, 
as expected, narrower than the original in all the energy range considered in this study. For an event energy of 1 TeV the 
new function is $\sim$4 times narrower than that corresponding to H.E.S.S. (Figure \ref{PSF}).

\vspace{.5cm}

\vspace{.5cm}
A narrower PSF means better angular resolution, which increases the telescope capability of distinguishing an extended component.

%COMENTARIO DEL AREA EFECTIVA TAL VEZ SUBESTIMADA???????%%%%%%%%%

\section{Method and results}
\label{met}

There are several possible ways for searching extended emission, most of them involve comparing the source angular distribution ($\theta^2$) with the one corresponding to a point source, i.e. the telescope PSF. Any significant difference between these two distributions would indicate the existence of an extended contribution. The ELMAG code generates the $\theta^2$ distribution of gamma rays when they reach the Earth, however, in order to compare it with the telescope PSF we need the $\theta^2$ distribution as seen by the telescope system. To do this we developed an algorithm that calculates the angular distribution of the averaged number of gamma rays per unit of $\theta^2$ detected by the telescopes $\left(\frac{d \langle N_{\gamma}\rangle}{d \theta^2}\right)$, for a given luminosity ($L$) and observation time ($T_{obs}$). This is achieved by sampling the effective area and the PSF for the future array calculated in section \ref{tel} and convolving it with the $\theta^2$ output of the simulation.

A Cherenkov telescope not only detects events from gamma rays but also from cosmic rays, which are the most important source of contamination, or background. In order to discriminate gamma-ray from cosmic-ray detections the background must be well understood and properly extracted from the overall signal. In this case the background rate ($b$) per unit of $\theta^2$ and energy is estimated by using the signal ($N_{\gamma}$) to noise ($N_{back}$) ratio equation,
\begin{equation}
\frac{N_{\gamma}}{\sqrt{N_{back}}}=\frac{\phi (E)\  A(E)\ \Delta E\ T_{obs}}%
{\sqrt{b(E)\ \psi_{68}^2\left(E\right)\ \Delta E \ T_{obs}}},
\end{equation}
where $E$ is the energy, $\phi(E)$ is the gamma-ray flux, $A(E)$ is the effective area, and $\Delta E$ is the energy bin width. The ability to discriminate background events together with the technical requirements to establish a source detection determine the sensitivity of a telescope system. By definition, if $N_{\gamma}/\sqrt{N_{back}}=5$, then the gamma-ray flux $\phi(E)$
corresponds to the sensitivity $S(E)/E^2$. Therefore, we have:
\begin{equation}
b(E) = \left[\frac{S\left(E\right)/E^2\ A\left(E\right)}{5\ \psi_{68}(E)}\right]^2\ \Delta E\ T_{obs}.
\label{back}
\end{equation}
The total background rate ($b_T$) is obtained combining Eqs. (\ref{sens}), (\ref{area}), (\ref{psi}) and (\ref{back}), 
and integrating in the considered energy range from 30 GeV to 10 TeV.

%We use a $\chi^2$ test, which is based on binned data, to compare the PSF with the observed $\theta^2$ distribution. 

Having the distribution of both the gamma rays from the source and the background rate, the next step is to construct the distributions corresponding to the ON and OFF source observations that allow to compute the significance of the simulated observation. To construct the ON distribution the number of events in the $i$-th $\theta^2$ bin is obtained by taking a random value from a Poisson distribution with mean $\mu\left(\theta^2 \right)_i = \frac{d \langle N_{\gamma}\rangle}{d \theta^2}(\theta^2_i) \times \Delta \theta^2$, where $\theta^2_i$ is the center of the bin and $\Delta \theta^2$ is the bin width. The background in each $\theta^2$ bin is calculated also by taking a random value from a Poisson distribution with mean $\mu\left(b\right) = b_T \times \Delta \theta^2$. Then, the ON distribution is obtained by adding the background to each $\theta^2$ bin.   
In a real observation, we consider that ON and OFF are taken from the same data set, i.e. wobble mode, in which the source is place off the center of the camera allowing to have more than one region in the sky where to compute the OFF \cite{daum1997}. 
We then construct a wobble background considering 3 such regions in the sky (OFF) by taking a random value from a Poisson distribution with mean $\mu\left(b\right) = 3 \times b_T \times \Delta \theta^2$.
Once obtained, this background is extracted from the ON distribution to get the excess distribution: ON-OFF/3. 

The comparison between the point source distribution (PSF) with the excess distribution was done using the $\chi^2$ test as follows:

\begin{equation}
\chi^2=\sum_{i}\frac{(N_{ex,i}-N_{PSF,i})^2}{N_{ex,i}+\frac{4}{3}N_{OFF,i}},
\end{equation}
where $N_{ex,i}$ is the excess and $N_{PSF,i}$ is the weighted sampled PSF in the $i$-th bin. The denominator contains the fluctuations of the quantities involved in the calculation, which follow Poisson statistics. $N_{OFF,i}$ is the fluctuation that comes from adding and subtracting the background (wobble) to obtain the excess. Note that the test is applied after both distributions are scaled so that the total number of events match to each other ($\sum_i N_{PSF,i}=\sum_i N_{ex,i}$).\\
 The $\chi^2$ test assumes Gaussian statistics for its variables, so a relatively large number of counts per bin is required. A common criteria is that up to 20\% of bins with $< 5$ counts is considered acceptable for the test to be used \cite{frodesen1979}. Given the nature of the distributions being compared, where most bins contents fluctuate around 0, the validity of the $\chi^2$ test is not obvious.  We therefore calculated the percentage of bins with less than 5 counts which in our case turn out to be significantly  lower than the requirement (i.e. $<1\%$). We also calculated the distribution of $\chi^2$ and compare it with the {\it chi}-square distribution for a fixed number of {\it d.o.f}. These distributions show no significant differences, suggesting that the {\it chi}-square distribution is a good approximation of our probability function.

%Note that $\sum_i N_{PSF,i}=\sum_i N_{ex,i}$.

\begin{figure}[!ht]
  \centering
  \includegraphics[width=0.9\textwidth]{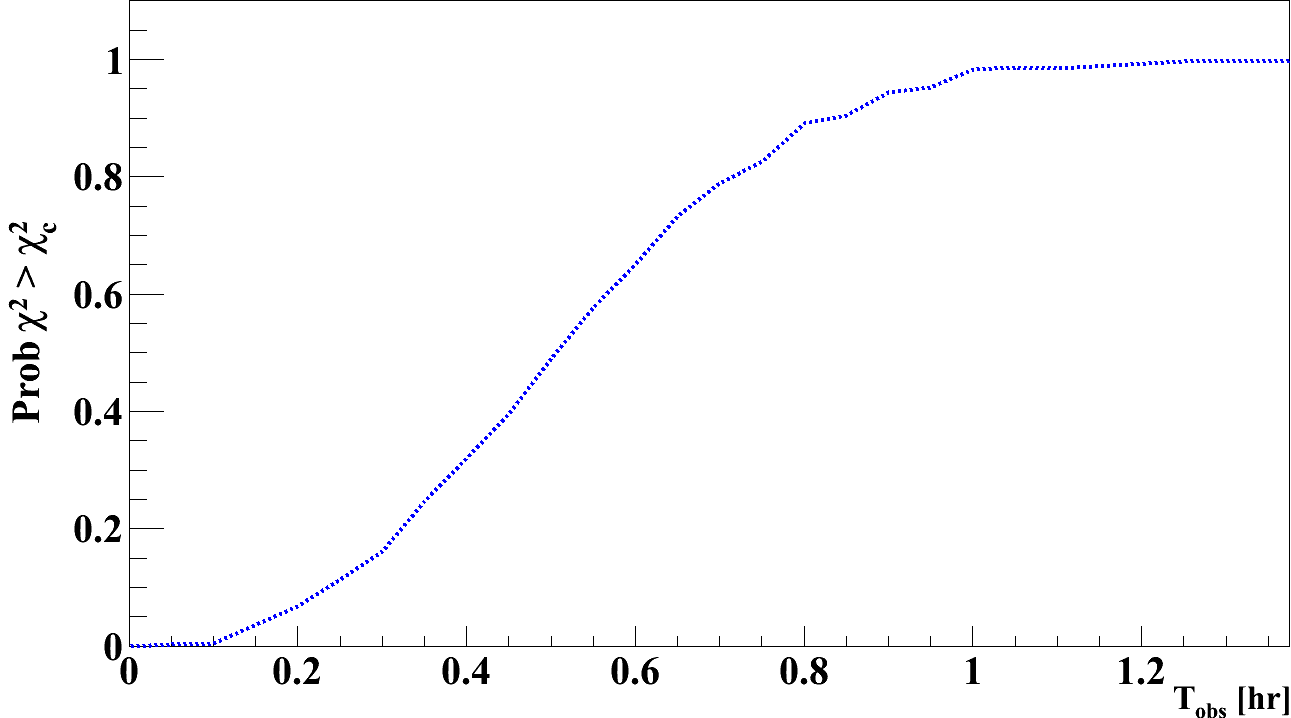}
  \caption{Fraction of samples that overpass the rejection threshold ($\chi_{c}$) as a function of the observation time.}
  \label{Prob}
\end{figure}

We take the non existence of an IGMF as the null hypothesis. For a fixed source luminosity and observation time, our code generates 1000 samples of the excess distribution and uses the test to reject the null hypothesis for a fraction of them. This fraction depends on the rejection probability, which in this work is taken as 99 \%. Figure \ref{Prob} shows the fraction of samples for which the null hypothesis is rejected as a function of the observation time T$_{obs}$. 
Although these results appear to be optimistic, one has to take into account that the models used to describe the telescope response are simple approximations that may be overestimating the capabilities of the telescopes.

%As expected the fraction increases with observation time, 
%saturating to one for luminosities larger than a given value. Results in figure \ref{Prob} also show that the fraction
%increases faster for longer observation times, reaching saturation at smaller luminosities.  
% 

\vspace{.5cm}

\section*{Conclusions}

The IGMF has a direct influence on the development of cascades from distant gamma-ray sources, which can be seen in both the spectral energy distribution and the angular ($\theta^2$) distribution. Unlike other previous studies that rely on differences in the SEDs to limit the IGMF intensity, our method focuses in the broadening effect that the magnetic field may produce in the angular distributions. In this way, the method allows to explore the chance of detecting extended emission for weaker IGMF intensities than other methods.
Combined with the ELMAG simulation, the method can be applied virtually to any VHE source in different redshifts and IGMF scenarios, offering a way to probe what the response of next generation Cherenkov telescopes would be, and what the chances are to detect these effects in the near future.

\end{document}